\documentclass[preprint,12pt]{elsarticle}
\usepackage[hidelinks]{hyperref} 
\usepackage{graphicx,import}
\usepackage{float}
\usepackage{caption}
\usepackage{subcaption}
\usepackage{setspace}
\usepackage{xcolor}
\usepackage{framed}
\usepackage{siunitx} 
\usepackage{ulem} 

\usepackage{amssymb} 
\usepackage{amsmath}
\usepackage[symbol]{footmisc}

\usepackage{lineno}
\usepackage{multicol}	

\journal{Elsevier}	

\begin{document}

\begin{spacing}{1.5}

\begin{frontmatter}

\title{Modelling wetting-bouncing transitions of droplet impact on random rough surfaces}


\author[a]{Huihuang Xia}
\author[b]{Yixiang Gan}
\author[a,c]{Wei Ge\corref{correspondingauthor}}
\cortext[correspondingauthor]{Corresponding author}
\ead{wge@ipe.ac.cn}

\address[a]{State Key Laboratory of Mesoscience and Engineering, Institute of Process Engineering, Chinese Academy of Sciences, Beijing 100190, China}
\address[b]{School of Civil Engineering, University of Sydney, NSW 2006, Australia}
\address[c]{School of Chemical Engineering, University of Chinese Academy of Sciences, Beijing 101408, China}

\begin{abstract}
Droplet impact on rough surfaces is of critical importance to various applications, yet remains incompletely understood. The present work aims to uncover droplet impact dynamics on random hydrophobic surfaces using volume of fluid simulations. Random fractal surfaces with RMS roughness ranging from 2 to 50 \si{\mu m} were generated using the Weierstrass–Mandelbrot function. Three identifiable impact outcomes including no bouncing, complete bouncing, and bouncing with breakup have been identified as Weber number varies between 5.7 and 12.9 and RMS roughness varies between 0 and 50 \si{\mu m}. We examine the spreading, retraction, re-spreading, and breakup stages of the impact events under different velocity and surface morphologies conditions. Numerical simulations show that the maximum spreading factor decreases linearly as surface roughness increases. Two scaling laws have been proposed for droplet impact on surfaces with small and large roughness values, respectively. A key finding is that the droplet contact time remains constant, independent of both Weber number and surface roughness. The joint effect of Weber number and surface roughness governs the wetting-bouncing transition, with larger roughness delaying the transition. This work elucidates the mechanisms governing droplet impact dynamics on random rough surfaces, thereby providing new insights directly relevant to droplet-based applications.
\end{abstract}

\begin{keyword}
Wetting transition \sep Droplet dynamics \sep Re-spreading \sep Rough surfaces \sep Contact time \sep Contact line
\end{keyword}

\end{frontmatter}


\section{Introduction}
\label{introduction} 
Droplet impact is ubiquitous in a wide range of applications, e.g., inkjet printing~\cite{derby2011inkjet}, spray cooling~\cite{jiang2022inhibiting}, coating~\cite{wu2022simulating}, integrated circuits~\cite{jiang2016fabrication} and binder jet 3D printing~\cite{wagner2024coupled}. Within the past decades, droplet impact dynamics have been investigated extensively, and different impact outcomes, for instance, droplet deposition, partial or total rebound, receding breakup, and splashing have been reported to elucidate the mechanisms of impact phenomena~\cite{josserand2016drop,gabbard2025drop,lathia2025hydrophobic,bao2023modified}. 
Droplet impact dynamics are governed by droplet properties (e.g., density, viscosity, and surface tension), surrounding air characteristics, impact parameters (e.g., impact velocity and angle) and surface properties that govern droplet wettability~\cite{josserand2016drop,richard2002contact,bird2013reducing,roisman2015drop}. 
Focus on surface morphologies as a research gap. For instance, a recent study has shown that wettability on rough surfaces affects the dynamic contact angle of droplets; namely, the maximum dynamic contact angle of an impacting droplet increases with an increase of the surface roughness at the micrometer scale~\cite{quetzeri2019effect}.  Gao et al. (2025) conducted two-dimensional simulations to understand how surface skewness and kurtosis affect droplet impact behaviour~\cite{gao2025numerical}. They pointed out that a droplet impacting on a rough surface with skewness = 0.2 and kurtosis = 3 shows superior rebound performance, and, notably, the overall droplet contact time on such a rough surface shows a reduction by $16.15\%$ compared to that on a smooth surface. 

Several dimensionless numbers, e.g., the Weber number $\textit{We}=\rho D_0U_0^2/\sigma$, the Reynolds number $\textit{Re}=\rho D_0U_0/\mu$, and the Ohnesorge number $\textit{Oh}=\mu/\sqrt{\rho D_0\sigma}$ have been adopted to describe droplet impact dynamics, where $U_0$ is the impact velocity, $D_0$ the droplet diameter, and $\rho$, $\mu$ and $\sigma$ are droplet density, dynamic viscosity and surface tension coefficient, respectively. Spreading dynamics of high-viscosity droplets have been found to be mainly affected by the Reynolds number \textit{Re}, as the viscous dissipation is dominant in this scenario~\cite{liu2024droplet}. An impacting low-viscosity droplet spreads and reaches its maximum spreading diameter $D_m$. Then it enters a phase of repeated receding and spreading until it finally sits on a substrate with an equilibrium contact angle $\theta_0$~\cite{bayer2006contact}. The maximum spreading diameter is of great importance to control the printing resolution and to minimize the void defects in inkjet printing technologies~\cite{derby2011inkjet}. A low-viscosity droplet impacting on a superhydrophobic surface has been investigated experimentally in the range of $2<\textit{We}<900$, and it shows that the maximum spreading diameter $D_m$ scales as $D_0\textit{We}^{1/4}$~\cite{clanet2004maximal}. A crossover regime between the viscous and capillary regimes has been proposed for droplet impacting on hydrophobic surfaces, and the scaling law $D_m/D_0 \sim \textit{Re}^{1/5}f(\textit{We}\textit{Re}^{-2/5})$ has been validated by experimental and numerical results~\cite{eggers2010drop,laan2014maximum}. Liu et al. (2025) derived $D_m/D_0 \sim (23.3 + \textit{We}/\textit{Oh})^{1/6}$ for droplet impacting on flat surfaces for $10^{-3} \leq \textit{Oh} \leq 10^0$ and $10^1 \leq \textit{We} \leq 10^3$, and this semi-empirical model shows good agreement with various experimental data~\cite{liu2025maximum}.

Contact time between an impacting droplet and a surface is significant for designing water-repellent, self-cleaning, and anti-icing surfaces~\cite{bird2013reducing,kreder2016desigh}. The contact time $\tau_c$ is independent of the impact velocity over the range of $0.2 - 2.3 \ \si{m/s}$, but $\tau_c$ has been reported to depend on the droplet radius $R$, namely, it scales as $\tau_c \sim \sqrt{\rho R^3/\sigma}$ for droplet impact on macroscopically flat superhydrophobic surfaces~\cite{richard2002contact}. Experimental studies have shown that superhydrophobic surfaces with microscopic texture significantly reduce the overall contact time of bouncing drops~\cite{bird2013reducing,gauthier2015water}. The underlying reason for the reduction in contact time is that surface morphology redistributes the liquid mass, thereby altering the impact dynamics of droplets. Similarly, Liu et al. (2014) designed a surface with tapered micro/nanotextures, and such a surface promotes a rapid droplet detachment and the occurrence of pancake bouncing~\cite{liu2014pancake}. Submillimeter droplets impacting on a rough surface with periodic microprotrusions have been investigated using Volume of Fluid (VoF) simulations, which show that the rough surface induces flow disturbances and enhances viscous dissipation during droplet spreading phase~\cite{yang2025impact}. Consequently, it takes longer for droplets to reach their maximum spreading diameter.

In addition to the aforementioned rough surfaces with tailored surface morphology, random rough surfaces have been adopted to investigate droplet impact dynamics. For instance, phase field simulations show that the evolution of droplet shape when impacting on two-dimensional random surfaces is quite different from the scenario when impacting on a patterned surface with the same surface roughness~\cite{xiao2018computational}. The root-mean-square (RMS) roughness has negligible influence on droplet splashing velocity when the roughness is minor ($< 3-4 \ \si{\mu m}$), and the splashing velocity decreases with increasing roughness when the roughness is larger than the above threshold~\cite{de2021droplet}. Bao et al. (2023) experimentally investigated the influence of the random surface roughness on the transition from droplet wetting to bouncing. They found that the wetting-bouncing transition can be triggered by increasing \textit{We}~\cite{bao2023modified}. Wang et al. (2024) investigated droplet dynamics when impacting on five random fractal surfaces using three-dimensional VoF simulations, and they demonstrated that a higher surface roughness produces a thicker liquid film on a rough  surface~\cite{wang2024numerical}. Tong et al. (2025) found that the critical \textit{We} required for a transition from droplet complete rebound to fragmentation splashing decreases monotonically with increasing surface roughness~\cite{tong2025mechanism}. Various roughness parameters have been utilized to study the effect of surface roughness on droplet impact dynamics, for instance, the root-mean-square roughness ($R_q$), the arithmetic average roughness ($R_a$), the maximum height of the profile ($R_z$), the average peak to peak feature size ($R_p$), and the mean width of the surface feature ($R_m$)~\cite{bao2023modified, quetzeri2019effect,wang2024numerical}. Roisman et al. (2015) found that the absolute length scales of the substrate roughness, e.g., $R_a$ and $R_z$,  have negligible influence on the splash threshold~\cite{roisman2015drop}.

In the aforementioned experimental and numerical investigations, droplet dynamics when impacting on smooth, patterned, tailored rough, and random rough surfaces have been extensively investigated. It is clear that surface morphology, micro-structures, and roughness affect droplet dynamics. However, a comprehensive understanding of how surface roughness influences contact line dynamics and thereby governs the wetting-bouncing transition remains elusive. To address this gap, this work employs numerical simulations to elucidate the underlying mechanisms. The numerical framework uncovers the internal flow dynamics and the temporal evolution of the triple contact line, offering insights that are complementary to experimental observations. This work is structured as follows: Section \ref{theoryAndSetup} presents the mathematical formulation and theoretical background. Section \ref{Results} details the numerical results for droplet impact on surfaces with varying roughness values. The main conclusions and an outlook for future work are summarized in Section \ref{conclusion}. The numerical setup and validation are presented in the \textbf{Appendix A} and \textbf{B}, respectively.

\section{Numerical framework and rough surface generation}
\label{theoryAndSetup}
\subsection{Mathematical formulation}
\label{mathFormulations}
An improved numerical framework for modelling surface-tension-dominant two-phase flow, presented in our previous work~\cite{xia2025modelling}, is employed to investigate droplet impact dynamics in this work. The geometric VoF approach is employed to capture the sharp gas-liquid interface during droplet impact by solving the following transport equation
\begin{equation} \label{conVoFEqn}
\frac{\partial \alpha_l}{\partial t}+\nabla \cdot  (\alpha_l \mathbf{U})= 0,
\end{equation}
where $\alpha_l$ is the volume fraction of the liquid phase, and $\mathbf{U}$ the velocity field. In this work, we assume that both the liquid and gas phases are incompressible and immiscible, which leads to the continuity equation
\begin{equation} \label{continuityEqnDivFree}
\nabla \cdot{\mathbf{U}}=0.
\end{equation}
The momentum equation of the isothermal, Newtonian two-phase flow is given by
\begin{equation} \label{momentumEqn}
\frac{\partial(\rho\mathbf{U})}{\partial t} + \nabla \cdot (\rho\mathbf{U}\mathbf{U}) = -\nabla p + \nabla \cdot [\mu (\nabla \mathbf{U} + (\nabla \mathbf{U})^{T})] + \rho \boldsymbol{g} + \mathbf{F}_{\sigma},
\end{equation}
where $p$ is the pressure, and $\mathbf{g}$ the acceleration of gravity. The superscript $T$ represents the transpose of a tensor. $\mathbf{F}_{\sigma}$ is the surface-tension force term. $\mathbf{F}_{\sigma}$ is calculated by
\begin{equation} \label{FsigmaModel}
\mathbf{F}_{\sigma} = \sigma k \nabla \alpha_l,
\end{equation}
where $\sigma$ is the surface tension coefficient, and $k$ is the interface curvature.
The geometric fitting approach developed by Jibben et al.~\cite{jibben2019paraboloid} is employed to calculate the interface curvature $k$. Two functions used for the geometric fitting in either two or three dimensions are given by
\begin{equation} \label{fitParaboloid}
\begin{aligned}
f(x,\ z) = a_1x^2+a_0x+z \quad \text{in 2D,}\\
f(x,\ y,\ z) = a_1x^2+a_0x+a_3y^2+a_2y+a_4xy+z \quad \text{in 3D.} \\
\end{aligned}
\end{equation}
In Eqn.~\ref{fitParaboloid}, $x$, $y$ and $z$ represent local rotated Cartesian coordinates with $z$ pointing along the interface normal, and $a_0$, $a_1$, $a_2$, $a_3$ and $a_4$ are calculated by a least square minimisation procedure. The interface curvature is then computed by
\begin{equation}
\begin{aligned}
k = \frac{f_{xx}}{(1+f_x^2)^{1.5}} \quad \text{in 2D,} \\
k = \frac{f_{xx}(1+f_x)+f_{yy}(1+f_y)-2f_xf_yf_{xy}}{(1+f_x^2+f_y^2)^{1.5}} \quad \text{in 3D.}
\end{aligned}
\end{equation}
An additional numerical filtering step is used to eliminate unphysical spurious currents, and the corresponding numerical details can be found in our previous work~\cite{xia2025modelling}. 

The contact angle model is of great importance to predict droplet spreading and recoiling behaviour during impact. In this work, we employ the contact angle model proposed by Seebergh et al.~\cite{seebergh1992dynamic}. The contact angle $\theta$ is given by
\begin{equation} \label{SeeberghModelEqn}
\theta =
\begin{cases}
\text{arcos}\left[\text{cos}\theta_e-4.47\textit{Ca}^{0.42}(1+\text{cos}\theta_e)\right], & \textit{Ca} \leq 0.001, \\
\text{arcos}\left[\text{cos}\theta_e-2.24\textit{Ca}^{0.54}(1+\text{cos}\theta_e)\right], & \textit{Ca} > 0.001,
\end{cases}
\end{equation}
where $\theta_e$ is the equilibrium contact angle, and \textit{Ca} the Capillary number ($\textit{Ca}=\mu U_0/\sigma$). Seebergh et al. (1992) pointed out that surface roughness induces a stick-slip effect in the low \textit{Ca} regime when the static contact angle is larger than around $15^{\circ}$. Note that Eqn.~\ref{SeeberghModelEqn} is still valid even in the presence of the above stick-slip effect.

\subsection{Generation of random rough surfaces}
\begin{figure}[htp]
  \begin{center}
    \includegraphics[width=0.9\textwidth]{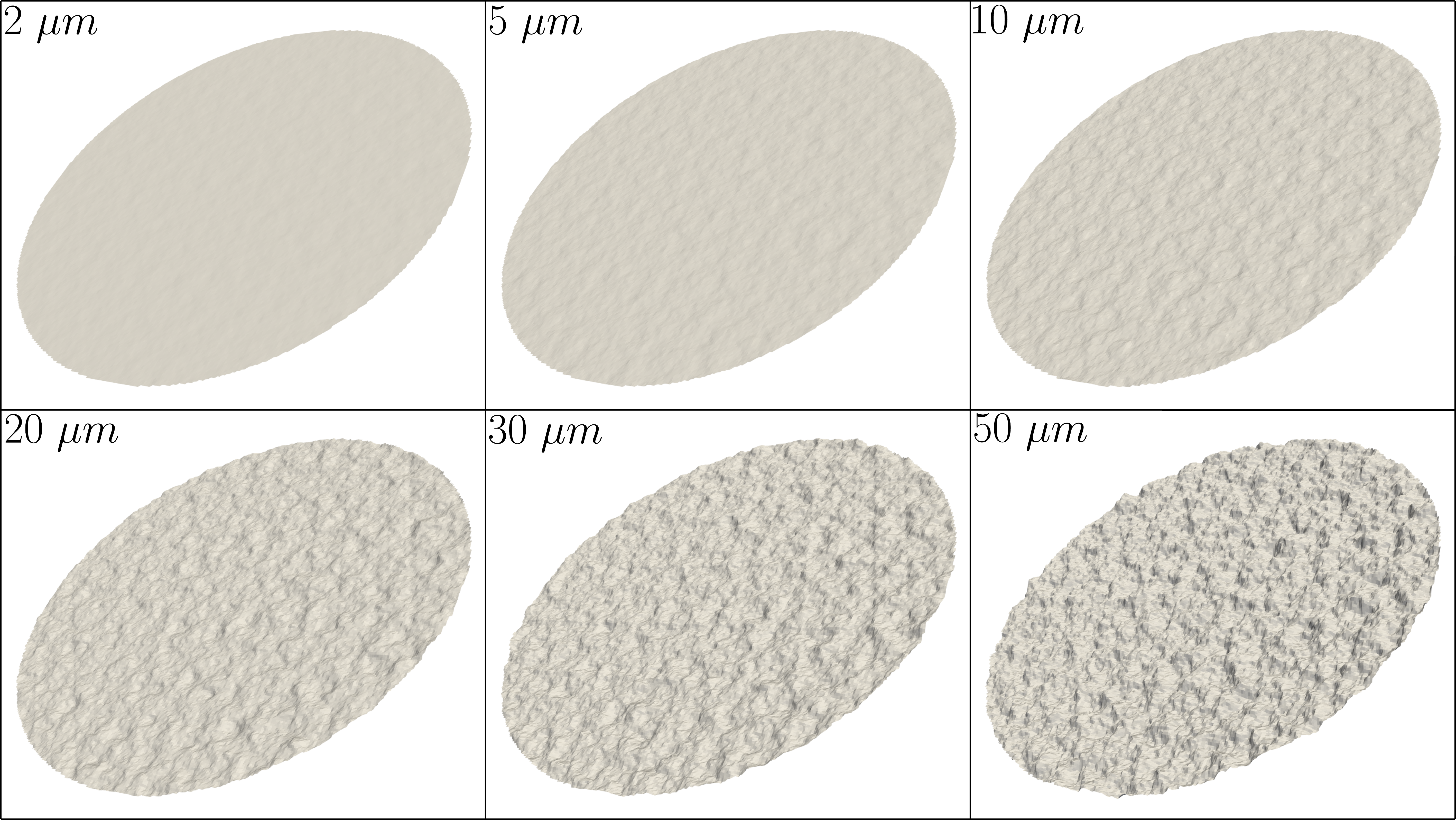}
  \end{center}
  \caption{Random rough fractal surfaces with varying root-mean-square roughness values generated using the W-M function.}
  \label{roughSurface}
\end{figure}

The Fast Fourier Transform (FFT) approach has been used to generate random rough surfaces with specified statistical parameters, e.g., skewness and kurtosis values~\cite{gao2025numerical,zhang2024dynamic}. However, random rough surfaces exhibit a form of self-similarity comparable to that of various intricate surfaces and phenomena in nature. Fractal theory has been proven to be capable of characterizing rough surfaces with fractal characteristics~\cite{wang2024numerical,barnsley2014fractals}. 
In this work, the Weierstrass–Mandelbrot (W-M) function is thus employed to generate fractal rough surfaces. The roughness height $z(x,y)$ at the spatial coordinates $(x,y)$ is given by
\begin{equation} \label{WMFunc1}
z(x,y) = A \sum_{m=1}^{M} \sum_{n=1}^{n_{\text{max}}} \gamma^{(D_f-3)n}\left[ \cos \phi_{mn} - \cos B \right],
\end{equation}
where $A$, $B$ is defined by
\begin{equation} \label{AandB}
\begin{split}
&A = L \left( \frac{G}{L} \right)^{D_f-2} \sqrt{\frac{\ln \gamma}{M}},\\
&B = 2\pi \gamma^n\frac{\sqrt{x^2 + y^2}}{L} \cos\left(\arctan\left( \frac{y}{x} \right) - \frac{\pi m}{M}\right) + \phi_{mn},
\end{split}
\end{equation}
respectively. In Eqn.~\ref{WMFunc1}, $D_f$ is the fractal dimension, $n_{\text{max}}$ the maximum frequency index, $M$ the number of angular directions, and $\phi_{mn} \sim \mathcal{U}(0, 2\pi)$ a uniformly distributed random phase. In Eqn.~\ref{AandB}, $G$ is the scaling coefficient, $L$ the sampling length, and $\gamma$ the frequency scaling factor~\cite{majhor2025multifractal}. Six random rough surfaces characterized by the root-mean-square roughness $R_q$, namely, $R_q = 2,\ 5,\ 10,\ 20,\ 30$, and $50 \ \si{\mu m}$ have been generated, as shown in Figure~\ref{roughSurface}. 

\subsection{Numerical configurations}
In addition to the numerical setup presented in \textbf{Appendix A}, we outline additional numerical configurations in this section. Five different impact velocity $U_0=$ 0.5, 0.535, 0.6, 0.7 and $0.75 \ \si{m/s}$ used in our simulations correspond to five different Weber numbers \textit{We}= 5.7, 6.5, 8.2, 11.2 and 12.9, respectively. Surfaces with seven different roughness values ranging from 0 to $50 \ \mu m$ are used to investigate the influence of surface morphology on droplet impact and bouncing dynamics at five different \textit{We}. Accordingly, thirty-five numerical cases are simulated in this work.

\section{Results and discussion}
\label{Results}
\subsection{Droplet impact outcomes on random rough surfaces}
\begin{figure}[htp]
  \begin{center}
    \includegraphics[width=0.9\textwidth]{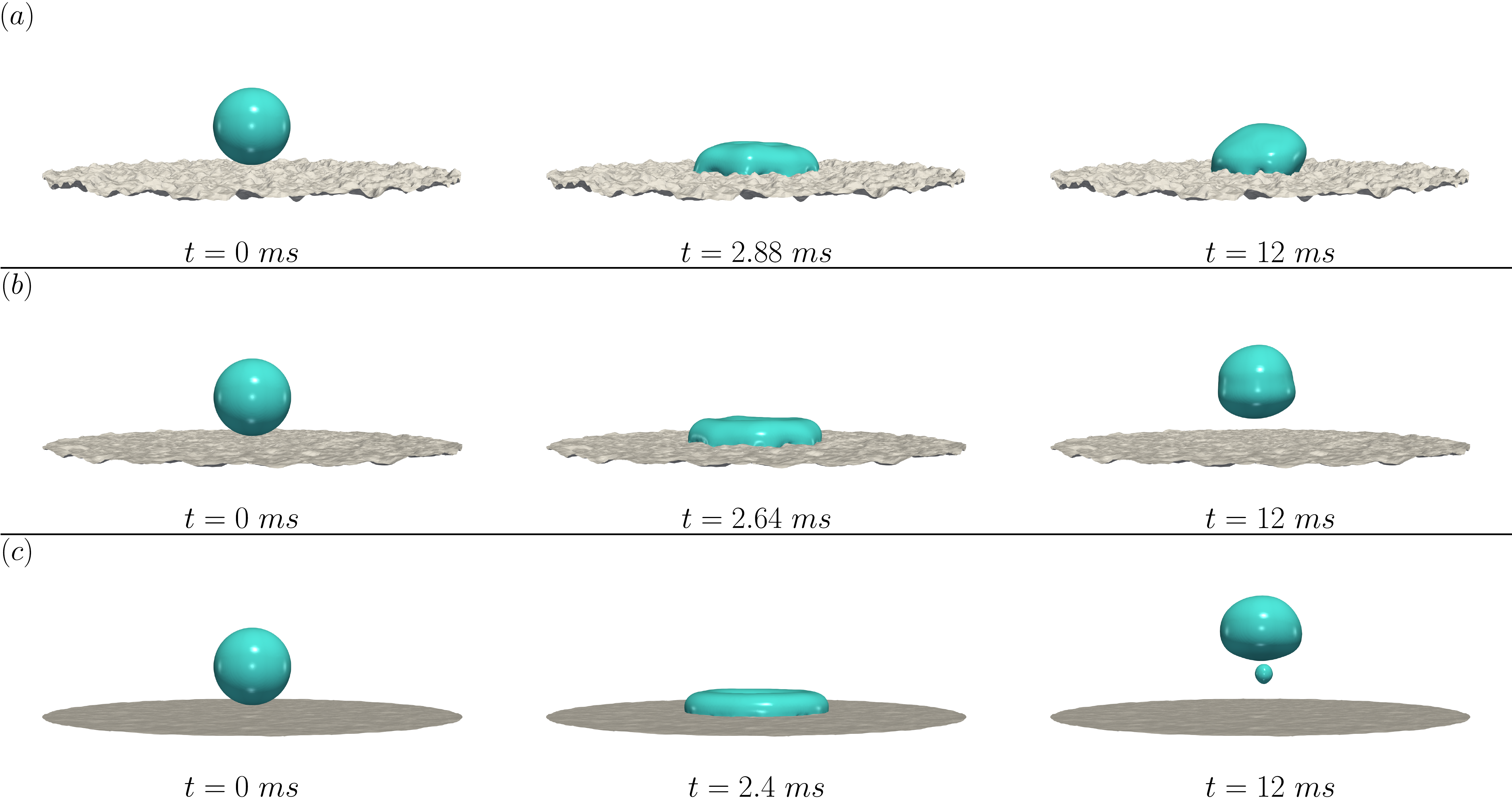}
  \end{center}
  \caption{Three different impact outcomes: (a)~no bouncing (\textit{We}=5.7, $R_q$=$50\  \mu m$), (b)~complete bouncing (\textit{We}=8.2, $R_q$=$20\ \mu m$), and (c)~bouncing with breakup (\textit{We}=12.9, $R_q$=$5\ \mu m$). The middle column shows droplet morphologies corresponding to the instant of the maximum spreading factor $\beta_m$.}
  \label{impactRegimes}
\end{figure}

Figure~\ref{impactRegimes} presents the temporal evolution of droplet deformation at three \textit{We}, and droplet reaches its maximum spreading factor $\beta_m$ at 2.88 \si{ms} for \textit{We} = 5.7, 2.64 \si{ms} for \textit{We} = 8.2, and 2.4 \si{ms} for \textit{We} = 12.9, respectively. The spreading factor $\beta$ is defined as the ratio of the projected wetting area on the horizontal base plane to the initial cross-sectional area ($\pi D_0^2/4$) of the spherical droplet before impact. The maximum spreading factor during droplet impact is denoted as $\beta_m$. Animations of the above three cases are available in the Supplementary Material.
Droplet impact on a rough surface with roughness value $R_q$ = 50 $\mu m$ at a low Weber number, i.e. \textit{We} = 5.7, ends up with droplet deposition, and we denote this impact outcome as \textit{no bouncing} in this work. With the increase of \textit{We}, the droplet rebounds completely from the rough surface with $R_q$ = 20 $\mu m$ at \textit{We} = 8.2, and this rebound is denoted as \textit{complete bouncing}.  A satellite droplet is observed (see Figure~\ref{impactRegimes}c) when the droplet rebounds from the rough surface with $R_q$ = 5 $\mu m$ at \textit{We} = 12.9, and this bouncing scenario is termed \textit{bouncing with breakup} or \textit{receding breakup}. Note that only three numerical cases are shown in Figure~\ref{impactRegimes}, representing three diverse impact outcomes. Thirty-two additional cases with varying \textit{We} and $R_q$ values exhibit the same impact scenario across the three aforementioned outcomes. The phase transition between wetting and bouncing, along with a more detailed analysis, will be discussed in the forthcoming sections.

\subsection{Droplet impact dynamics}
\begin{figure}[htp]
\centering
\begin{subfigure}[h]{0.495\textwidth}
    \includegraphics[width=\textwidth]{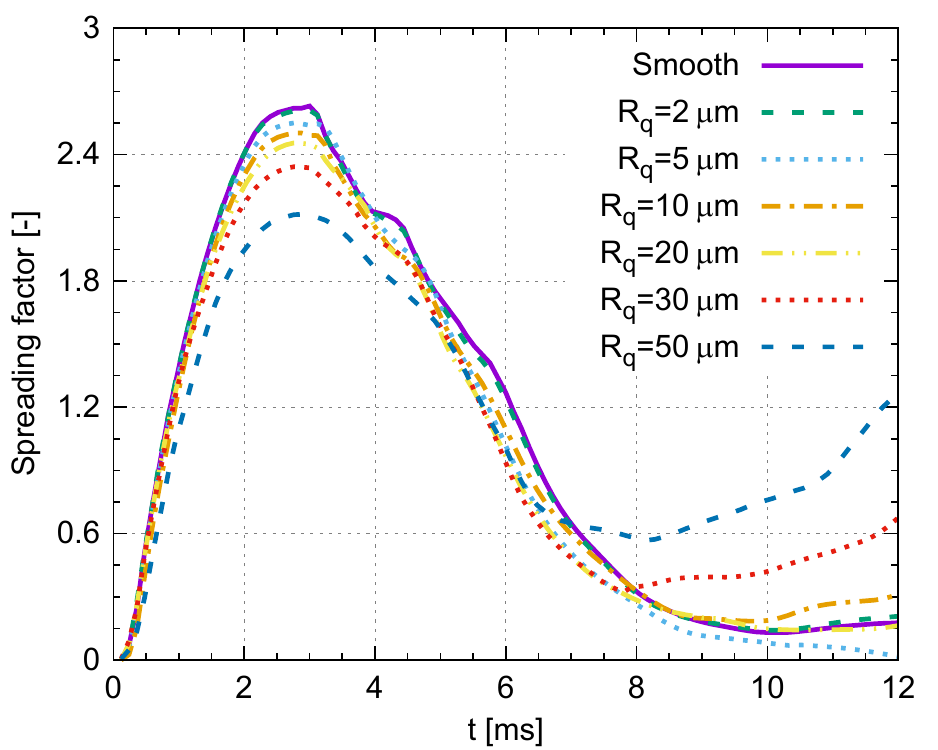}
    \caption{}
    \label{We_5.7_spreadF}
\end{subfigure}
\hfill
\begin{subfigure}[h]{0.495\textwidth}
    \includegraphics[width=\textwidth]{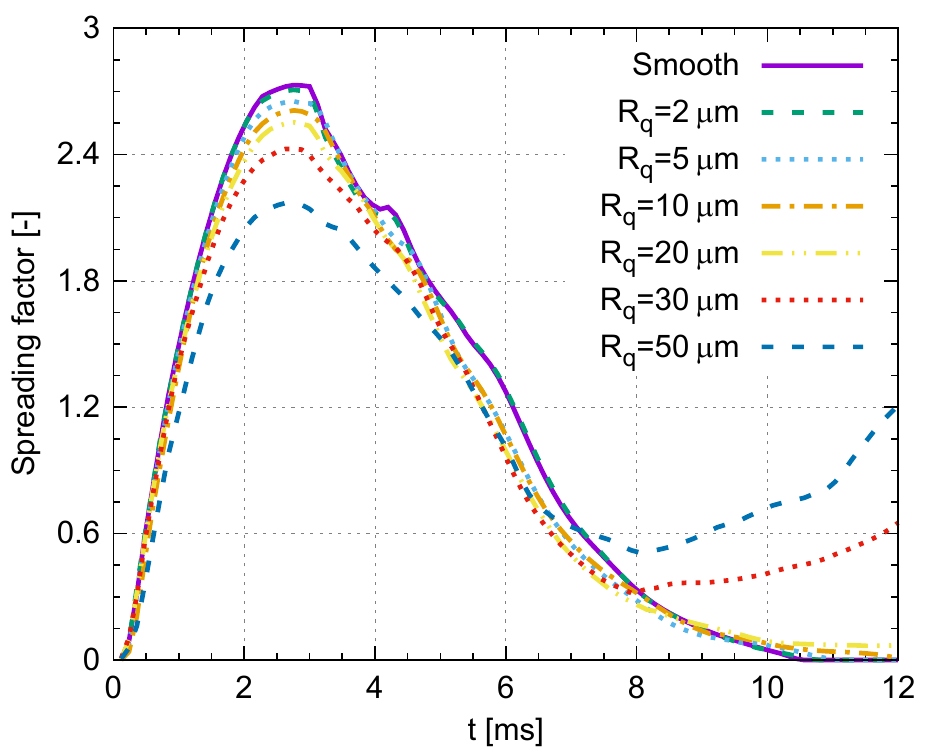} 
    \caption{}
    \label{We_6.5_spreadF}
\end{subfigure}

\begin{subfigure}[h]{0.495\textwidth}
    \includegraphics[width=\textwidth]{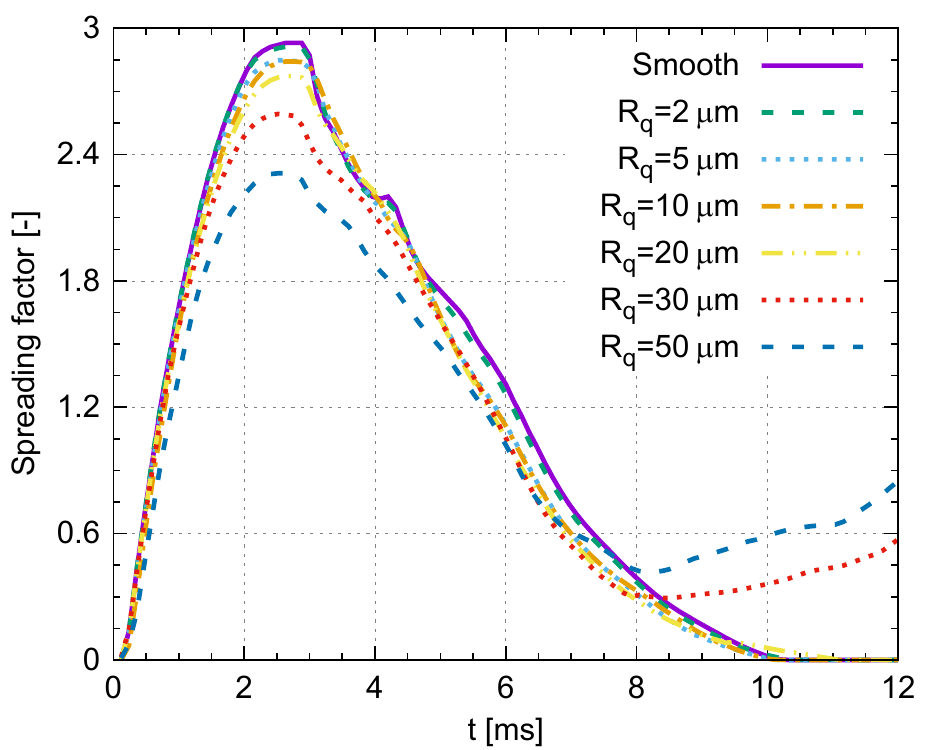}
    \caption{}
    \label{We_8.2_spreadF}
\end{subfigure}
\hfill
\begin{subfigure}[h]{0.495\textwidth}
    \includegraphics[width=\textwidth]{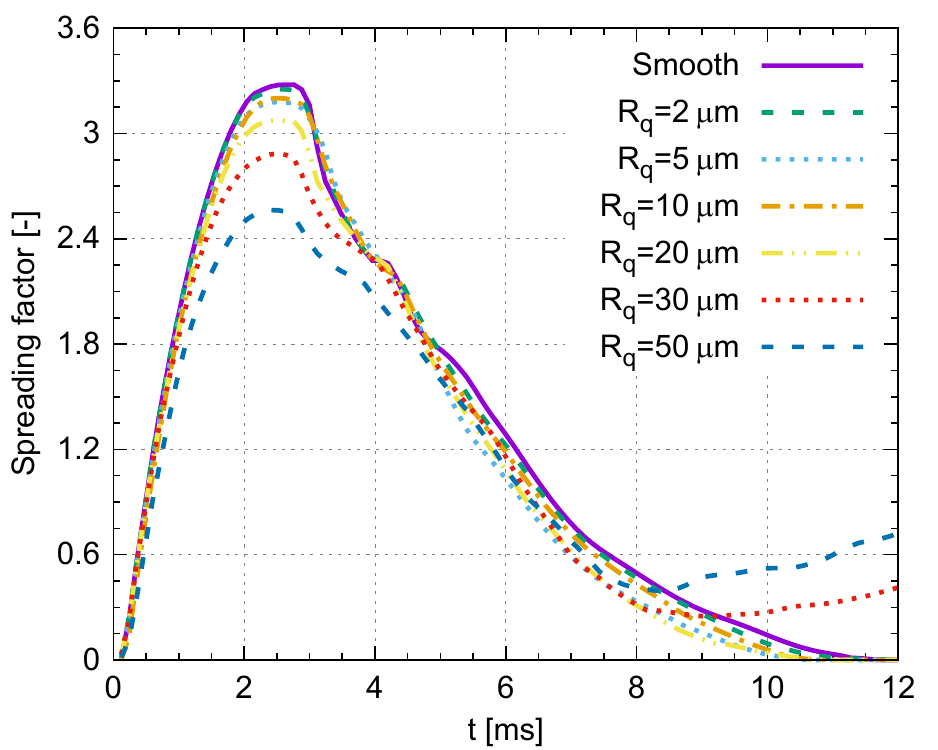}
    \caption{}
    \label{We_11.2_spreadF}
\end{subfigure}
\begin{subfigure}[h]{0.495\textwidth}
    \includegraphics[width=\textwidth]{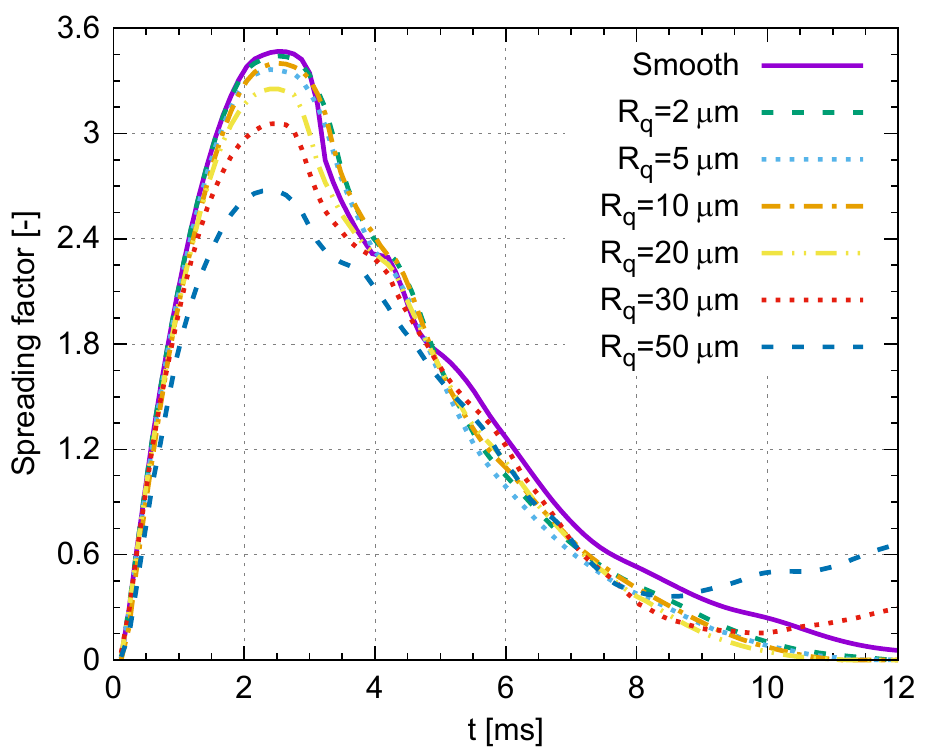}
    \caption{}
    \label{We_12.9_spreadF}
\end{subfigure}
\hfill
\begin{subfigure}[h]{0.495\textwidth}
    \includegraphics[width=\textwidth]{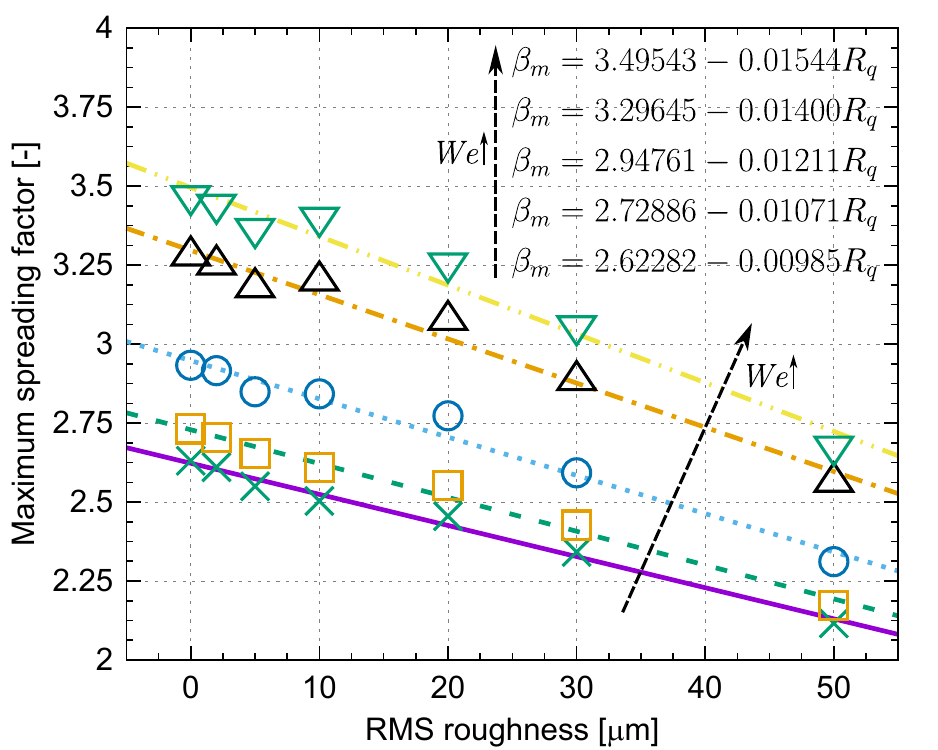}
    \caption{}
    \label{maxSpreadFMerge}
\end{subfigure}
\caption{Droplet impact dynamics at different \textit{We}. (a)-(e):~spreading factor versus time. (a)~\textit{We}=5.7, (b)~\textit{We}=6.5, (c)~\textit{We}=8.2, (d)~\textit{We}=11.2, (e)~\textit{We}=12.9, (f)~maximum spreading factor versus the RMS roughness.}
\label{maxSpreadF}
\end{figure}

The impact dynamics during droplet spreading and receding phases are discussed in this section. For each given Weber number, we investigate the influence of surface roughness on the droplet spreading factor. As shown in Figure~\ref{maxSpreadF}, we present the temporal evolution of the spreading factor up to 12 \si{ms}. 
Spreading factors of seven cases with different surface roughness at a given \textit{We} gradually increase, and then reach their maximum spearing factors $\beta_m$ during the spreading phase. Furthermore, a larger roughness value leads to a slower increase in the spreading factor during the spreading phase. At \textit{We}=5.7, non-zero spreading factors are observed when $t=$ 12 \si{ms} (see Figure~\ref{We_5.7_spreadF}). Accordingly, droplets deposit on surfaces after their recoiling phase. As \textit{We} continues to increase, more kinetic energy is transferred from the vertical to the radial direction, and thus results in larger spreading factors. Besides, spreading factors gradually decrease and reach zero for some droplets during their receding phase, as shown in Figure~\ref{We_8.2_spreadF}. This scenario evidences that droplets completely rebound from surfaces. 

Interestingly, droplets show non-zero spreading factors at $t=$ 12 \si{ms} when impacting on rough surfaces with roughness values $R_q \geq 30 \ \mu m$, across all five Weber numbers studied in this work. Furthermore, re-spreading behaviour occurs during the recoiling phase when impacting on rough surfaces with $R_q \geq 30 \ \mu m$. As pointed out by Luo et al. (2023), the threshold for the occurrence of re-spreading when impacting on smooth superhydrophobic surfaces is 4.9~\cite{luo2023re}. When \textit{We} increases from 5.7 to 12.9 for droplets impacting on hydrophobic surfaces with $R_q \geq 30 \ \mu m$, spreading factors at $t=$ 12 \si{ms} gradually decrease. For a given \textit{We}, the spreading factor during the re-spreading phase increases as the roughness value increases from 30 to $50 \ \mu m$. Accordingly, in this work, we note that the threshold for triggering re-spreading depends on both Weber number and surface roughness. 

As shown in Figure~\ref{maxSpreadFMerge}, solid or dashed lines represent the linear fitting of the maximum spreading factor $\beta_m$ represented by scattered symbols when impacting on surfaces with different roughness values $R_q$. Note that the ordinary least-squares method is used for data fitting. It is observed that $\beta_m$ decreases linearly with the increase of $R_q$ for each given \textit{We}. The linear fit yields the relationship $\beta_m = 2.62282 - 0.00985 R_q$ for \textit{We}=5.7. Similarly, linear relationships between $\beta_m$ and $R_q$ at other Weber numbers, namely, \textit{We} = 6.5, 8.2, 11.2, and 12.9, can also be obtained via linear fitting. The coefficients of determination $R^2$ for the above linear fits are all greater than 0.99. Additionally, it is noted that a higher \textit{We} yields not only a larger intercept but also a more rapid decrease of $\beta_m$.

\begin{figure}[htp]
  \begin{center}
    \includegraphics[width=0.5\textwidth]{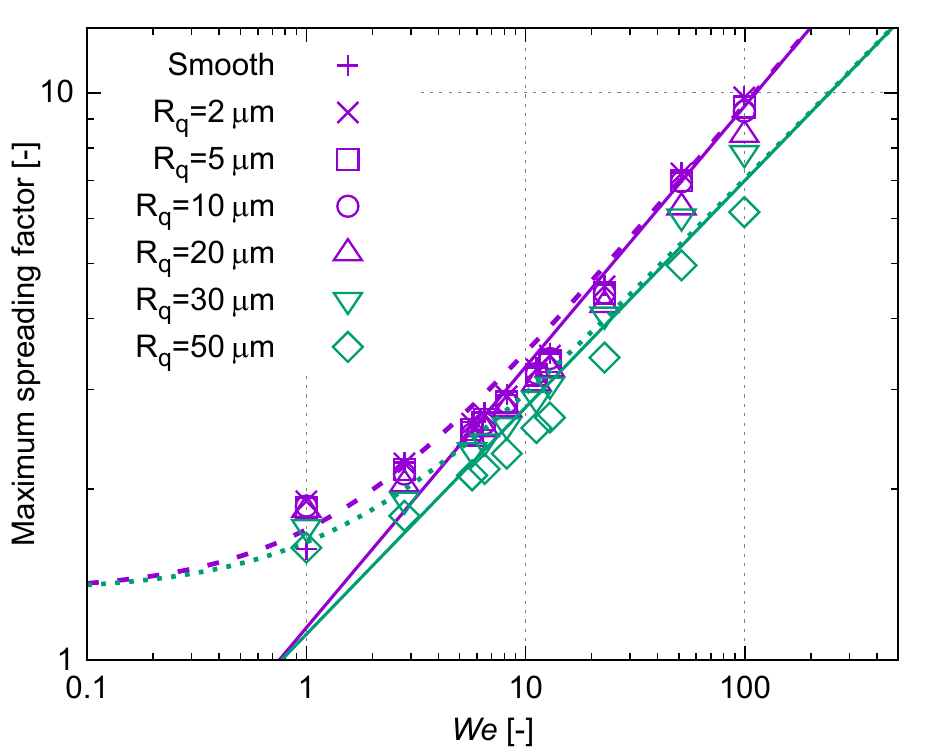}
  \end{center}
  \caption{Maximum spreading factor $\beta_m$ as a function of \textit{We}. Solid purple and green lines represent $\beta_m=1.14\textit{We}^{0.46}$ and $\beta_m=1.11\textit{We}^{0.40}$, respectively. Dashed purple and green lines represent $\beta_m=1.14(1.38+\textit{We})^{0.46}$ and $\beta_m=1.11(1.55+\textit{We})^{0.40}$, respectively.}
  \label{scalingLaw_We}
\end{figure}

To find the correlation between the maximum spreading factor $\beta_m$ and Weber number \textit{We}, additional numerical simulations with \textit{We}=1.0, 2.8, 22.9, 51.4, and 99.4 have been conducted. Note that such a wide range of Weber numbers is only used in this section to find a valid scaling law. The main objective of this work is to investigate the wetting-bouncing transition at low Weber number ($5.7 \leq \textit{We} \leq 12.9$). As shown in Figure~\ref{scalingLaw_We}, $\beta_m$ is found to increase as \textit{We} increases when \textit{We} varies between 1.0 and 99.4.
Figure~\ref{scalingLaw_We} demonstrates two scaling regimes governed by surface roughness values. For surfaces with lower roughness ($R_q \leq 20 \ \mu m$), $\beta_m$ follows $\beta_m = 1.14 \textit{We}^{0.46}$. $\beta_m = 1.11 \textit{We}^{0.40}$ is proposed for surfaces with higher roughness ($R_q \geq 30 \ \mu m$). However, the above scaling laws show underestimation of $\beta_m$, especially when \textit{We} is smaller than 5.7. As inspired by the semi-empirical scaling model proposed by Liu et al.~\cite{liu2025maximum}, a modified scaling model has thus been proposed, namely, $\beta_m \sim (A+\textit{We})^{0.46}$ with $A$ being a constant to be determined. In practice, a droplet gradually spreads on a surface and finally sits on this surface with an equilibrium contact angle when depositing such a droplet with a very small velocity.  $A=1.38$ and $A=1.55$ can be obtained by applying $\lim\limits_{\textit{We} \to 0}\beta_m$=1.322 for $\beta_m \sim (A+\textit{We})^{0.46}$ and $\beta_m \sim (A+\textit{We})^{0.40}$, respectively. As shown in Figure~\ref{scalingLaw_We}, modified scaling models show better performance in predicting $\beta_m$ when \textit{We} varies between 1.0 and 99.4. The reduction in the exponent from 0.46 to 0.40 evidences that higher roughness values hinder the inertial spreading of droplets. The above scaling laws quantify the transition from an inertia-dominated spreading regime on smooth surfaces to a roughness-inhibited regime where surface dissipation mechanisms significantly affect the spreading and deformation of impacting droplets. Additionally, $\beta_m$ as a function of $\textit{We}/\textit{Oh}$ also demonstrates two diverse regimes related to roughness values. Detailed information can be found in the \textbf{Appendix C}.

\begin{figure}[htp]
  \begin{center}
    \includegraphics[width=0.5\textwidth]{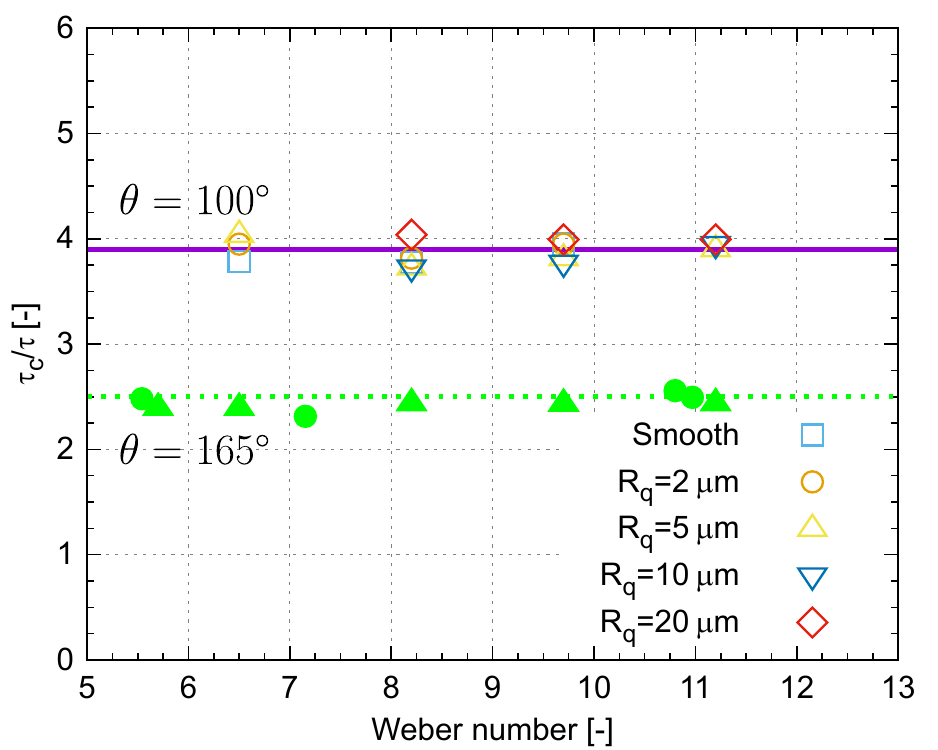}
  \end{center}
  \caption{$\tau_c/\tau$ as a function of Weber number. Open symbols: simulations for a contact angle $\theta=100^{\circ}$, solid triangle: present simulations for a contact angle $\theta=165^{\circ}$, solid circle: experimental results for a contact angle $\theta=165^{\circ}~\cite{gauthier2015water}$. Purple solid line and green dashed line represent $\tau_c/\tau$=3.9 and 2.5, respectively.}
  \label{contactTime}
\end{figure}

The duration a droplet stays in contact with a solid surface during impact was reported by Richard et al.~\cite{richard2002contact}. The contact time $\tau_c$ scales as $\tau_c \sim \tau = \sqrt{{\rho {R}^3}/{\sigma}}$ with $\rho$ and $R$ being droplet density and radius, respectively. Richard et al. (2002) pointed out that $\tau_c/\tau$ is $2.6 \pm 0.1$ when impacting on superhydrophobic surfaces. Both experimental~\cite{gauthier2015water} and numerical investigations~\cite{moqaddam2017drops} have evidenced the validity of the aforementioned scaling law. In addition to the above simulations on surfaces with a contact angle $\theta=100^{\circ}$, we conducted additional numerical simulations on superhydrophobic surfaces (either smooth or rough) with a contact angle $\theta=165^{\circ}$. As shown in Figure~\ref{contactTime},
solid triangles represent $\tau_c/\tau$ of simulations impacting on superhydrophobic surfaces. Solid circles show experimental results of $\tau_c/\tau$ on flat superhydrophobic surfaces adopted from the literature~\cite{gauthier2015water}. The green dashed line represents a ratio of $\tau_c/\tau$=2.5. This dashed line intersects all solid symbols, confirming the accuracy of our numerical model in predicting droplet contact time on superhydrophobic surfaces. Additionally, the open symbols shown in Figure~\ref{contactTime} illustrate numerical simulations on hydrophobic surfaces with a contact angle $\theta=100^{\circ}$. Remarkably, droplet contact time shows independence of both Weber number and surface roughness when $R_q \leq 20 \ \mu m$. Contact time on hydrophobic surfaces is longer than that on superhydrophobic surfaces, and $\tau_c/\tau$=3.9 is deduced from Figure~\ref{contactTime}. It indicates that the droplet contact time $\tau_c$ on hydrophobic surfaces (either smooth or slightly rough) can be predicted by $3.9 \sqrt{{\rho {R}^3}/{\sigma}}$.

\subsection{Transition from wetting to bouncing}
\begin{figure}[htp]
\centering
  \begin{subfigure}[h]{0.5\textwidth}
    \includegraphics[width=\textwidth]{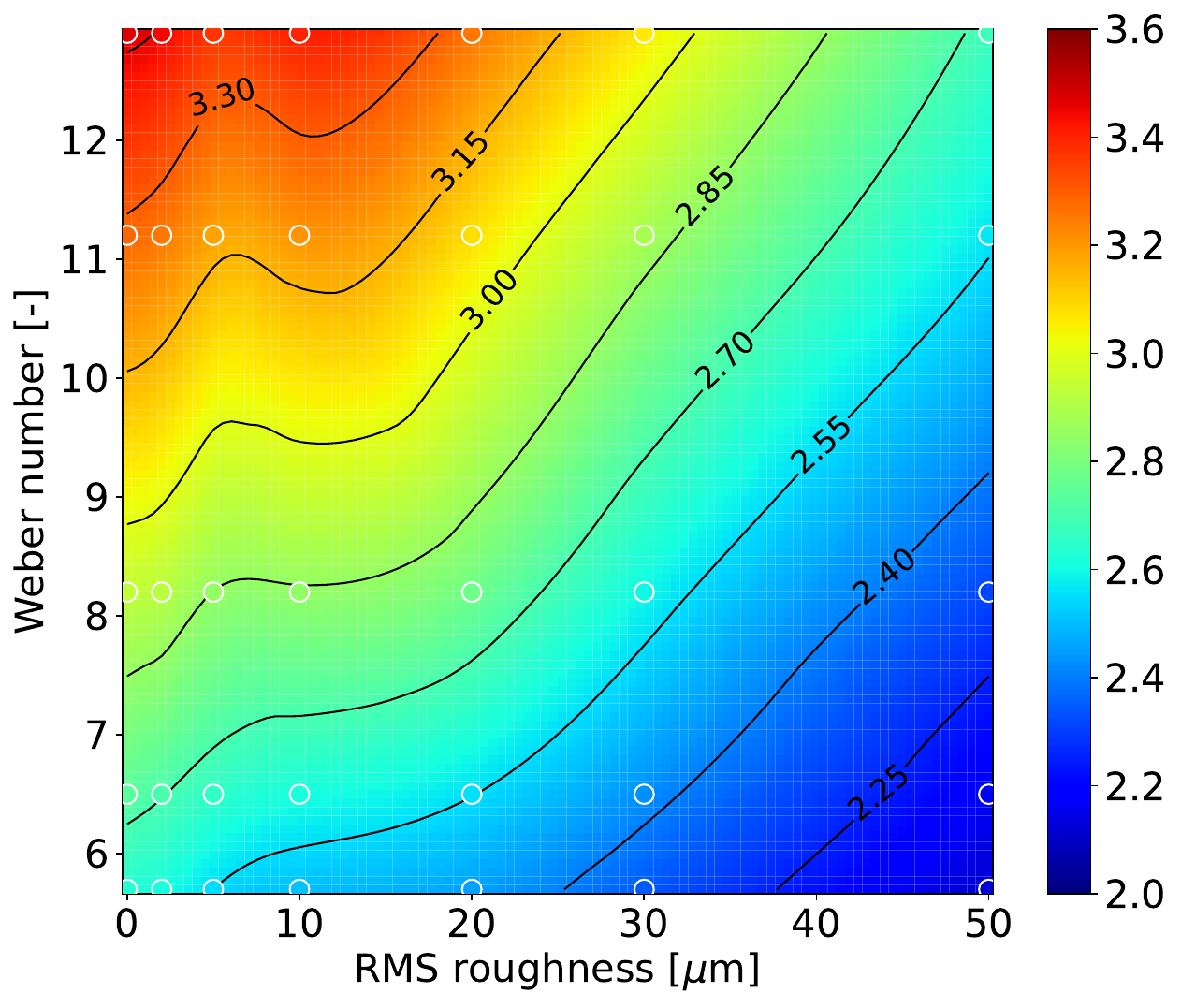}
    \caption{}
    \label{pD_maxSpFact}
  \end{subfigure}
  \hfill
  \begin{subfigure}[h]{0.425\textwidth}
    \includegraphics[width=\textwidth]{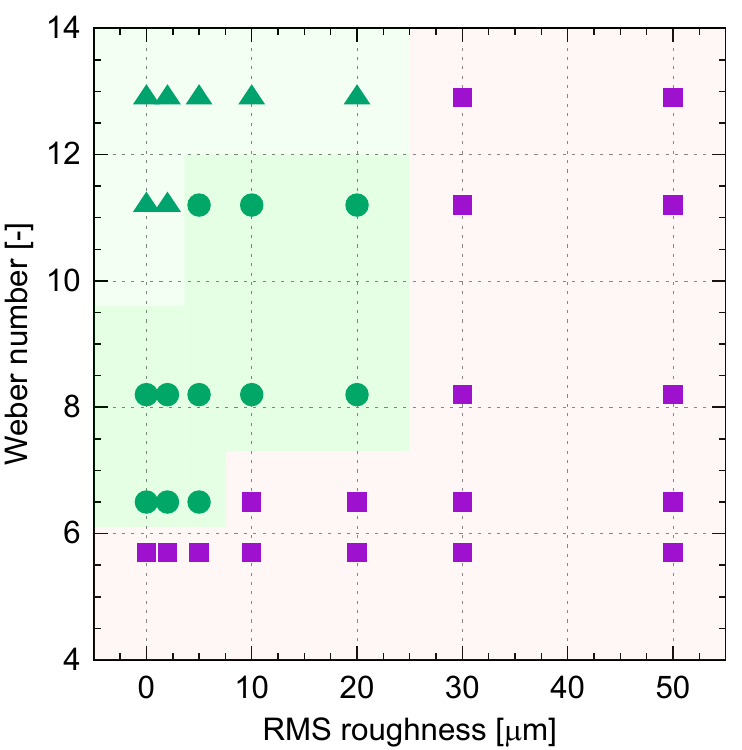}
    \caption{}
    \label{pD_bouncing}
  \end{subfigure}
  \caption{Droplet impact on rough surfaces with varying roughness values at different Weber numbers: (a)~contour plot of the maximum spreading diameter $\beta_m$, (b)~phase diagram for wetting-bouncing transition (purple square: no bouncing, green circle: complete bouncing, green triangle: bouncing with breakup).}
  \hfill
\end{figure}

This section discusses the effect of surface roughness on droplet bouncing dynamics and the wetting-bouncing transition. The maximum spreading factors $\beta_m$ on surfaces with different roughness values $R_q$ at different Weber numbers are shown in Figure~\ref{pD_maxSpFact}. 
The largest $\beta_m$, approximately 3.47, is observed for droplet impact on a smooth surface at $\textit{We}=12.9$. Droplet impact on the roughest surface with $R_q=$ 50 $\mu m$ yields the smallest spreading (approximately 2.12) at $\textit{We}=5.7$. The primary trend reveals that $\beta_m$ decreases as $R_q$ increases for a given $\textit{We}$. This decline becomes steeper as the roughness value increases. For instance, at $\textit{We}=$ 5.7, the value of $\beta_m$ drops from approximately 2.63 when $R_q$=0 to around 2.12 when $R_q$=50 $\mu m$. Simultaneously, $\beta_m$ exhibits a monotonic increase with \textit{We}. A higher \textit{We} consistently yields a larger $\beta_m$ for any fixed $R_q$. This effect is most evident at low roughness values, where $\beta_m$ increases from around 2.63 at $\textit{We}=$5.7 to approximately 3.47 at $\textit{We}=$12.9. Note that extreme surface roughness becomes the dominant factor, overwhelming the influence of the Weber number. This creates a distinct region in the high-roughness, mid-to-low \textit{We} area, characterized by the lowest values on the contour plot. Conversely, the region of low $R_q$ combined with high $\textit{We}$ exhibits the highest values, forming the opposite extreme on the contour plot. Figure~\ref{pD_maxSpFact} illustrates the combined influence of surface roughness and Weber number on droplet spreading dynamics. It highlights two distinct regimes, namely, a high-$\beta_m$ regime for high-inertia impacts on smooth surfaces and a low-$\beta_m$ regime on rough surfaces.

Figure~\ref{pD_bouncing} illustrates the phase diagram depicting the transition from wetting to bouncing for thirty-five cases simulated in this study. The purple square, green circle and green triangle in Figure~\ref{pD_bouncing} represent no bouncing, complete bouncing and bouncing with breakup, respectively (see Figure~\ref{impactRegimes}). The \textit{no bouncing} regime occupies the largest region of the phase diagram. It is characterised by relatively low Weber numbers ($\textit{We} \leq 8.2$) across a wide roughness range, namely, $0 \leq R_q \leq 50$. Notably, for the smallest Weber number of 5.7, \textit{no bouncing} occurs even on perfectly smooth surfaces ($R_q$ = 0), indicating that inertia is insufficient to promote droplet rebound at such a low \textit{We}. Furthermore, at a constant, intermediate Weber number (e.g., \textit{We} = 6.5 or 8.2), a transition from \textit{no bouncing} to \textit{complete bouncing} is observed as $R_q$ exceeds a critical threshold, approximately 25 for $\textit{We}$ = 8.2. This demonstrates the dominant role of increasing surface roughness in dissipating kinetic energy and suppressing the recovery of droplet shape. The \textit{complete bouncing} regime is confined to a band of moderate Weber numbers ($6.5 \leq \textit{We} \leq 11.2$) and low-to-moderate surface roughness ($R_q \leq 20$). In this region, droplet inertia is sufficient to overcome adhesion and viscous dissipation, but not so high as to induce interfacial breakup. The onset of breakup defines the upper boundary of this regime, while increased roughness values suppress its lower boundary. The \textit{bouncing with breakup} regime emerges at high Weber numbers ($\textit{We} \geq 11.2$) and on surfaces with low roughness ($R_q \leq 20$). At a given high \textit{We} (e.g., \textit{We}=11.2), increasing surface roughness appears to stabilise the droplet, and thus leads to a transition from \textit{bouncing with breakup} to \textit{complete bouncing}. It implies that \textit{bouncing with breakup} requires two concurrent conditions. First, a moderately high Weber number to induce a rapid and inertia-dominated retraction. Second, a smooth or rough surface with limited roughness values to minimize energy dissipation and facilitate the detachment of the rebounding droplet.

\subsection{Microscopic insights}
\subsubsection{Contact line dynamics}
\begin{figure}[htp]
\centering
  \begin{subfigure}[h]{0.495\textwidth}
    \includegraphics[width=\textwidth]{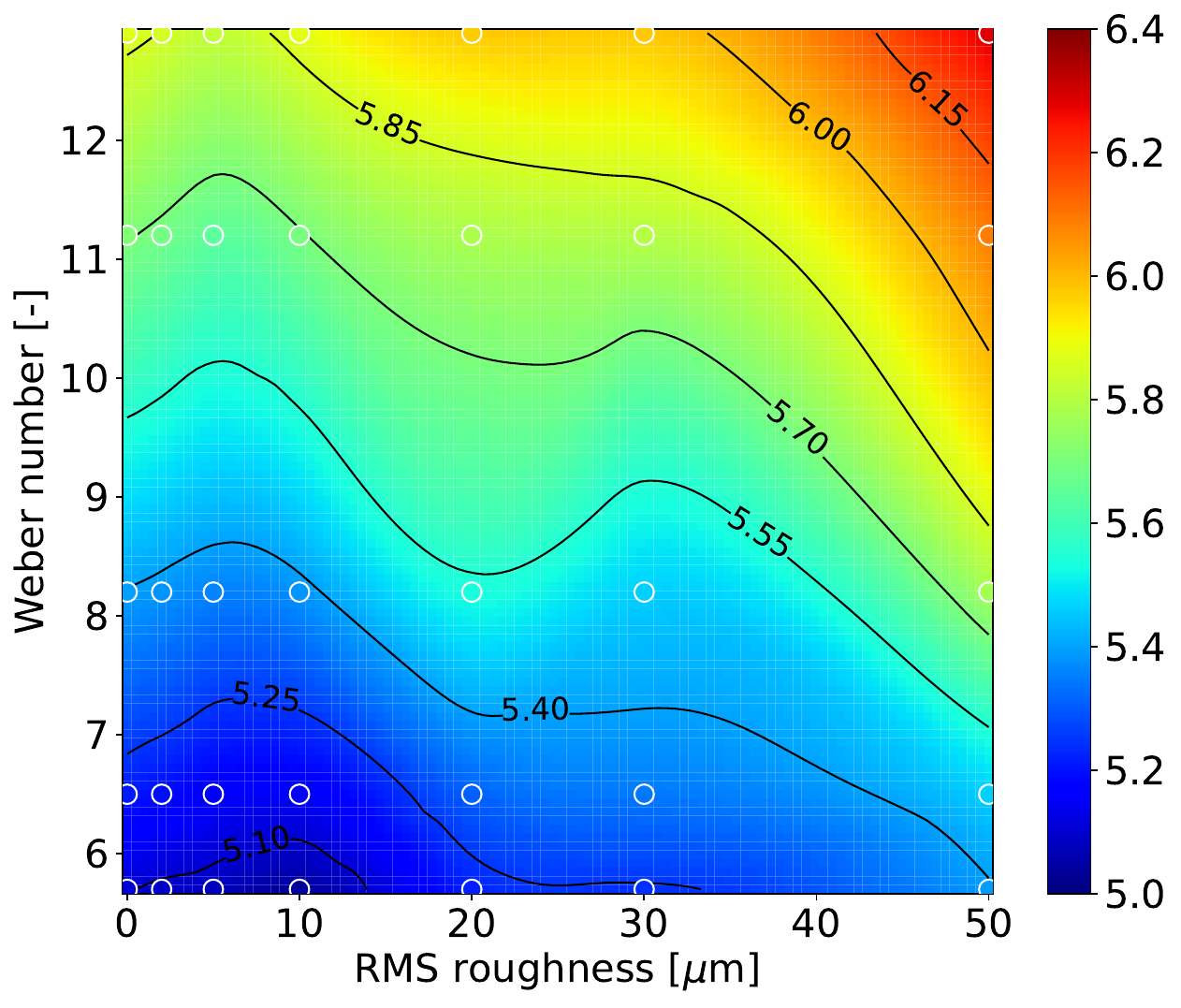}
    \caption{}
    \label{pD_CLLenngth}
  \end{subfigure}
  \hfill
  \begin{subfigure}[h]{0.495\textwidth}
    \includegraphics[width=\textwidth]{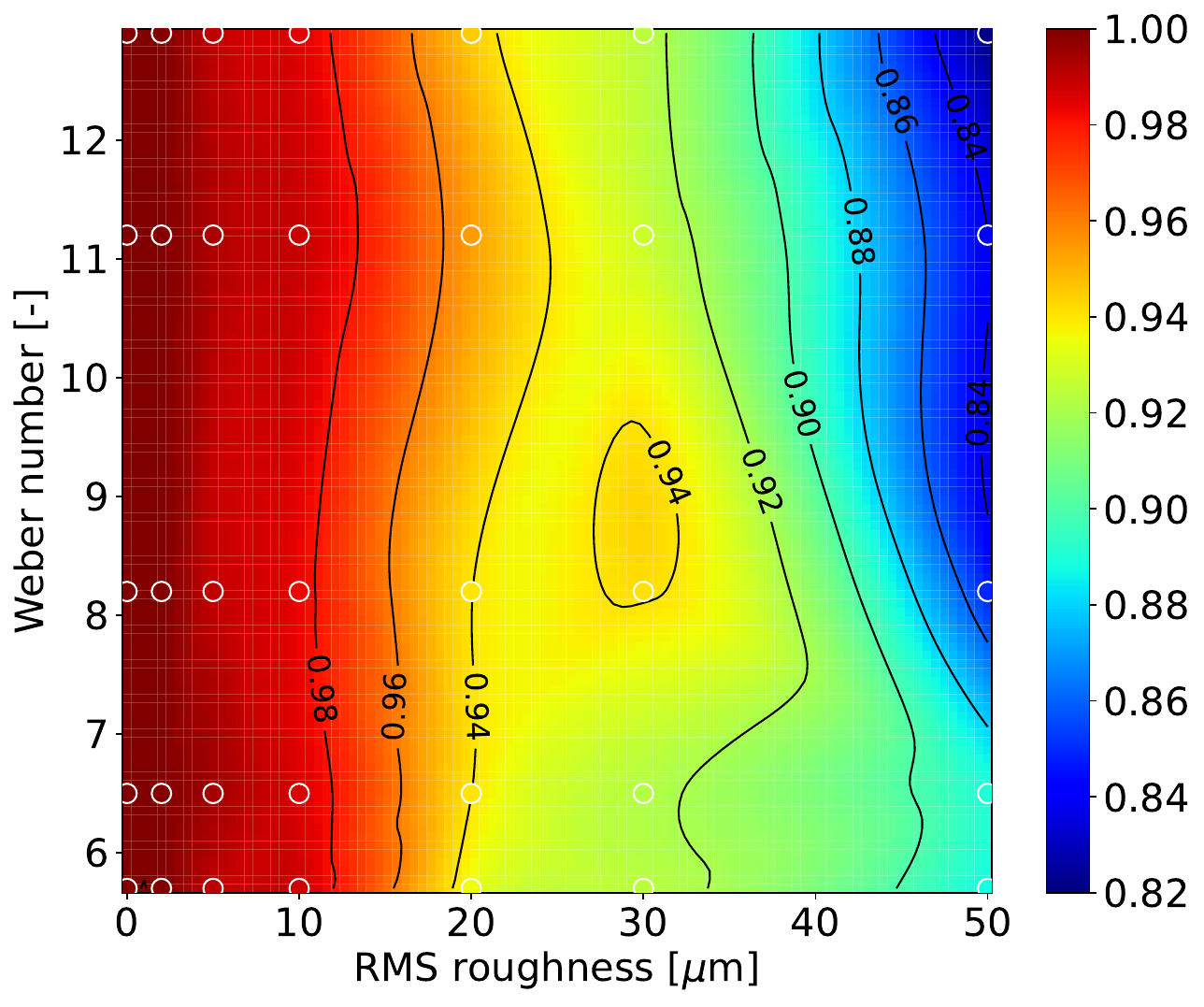}
    \caption{}
    \label{pD_circularity}
  \end{subfigure}  
  \caption{Contour plots of droplets impacting on rough surfaces with varying roughness values at different Weber numbers: (a)~dimensionless triple contact line length $\tilde{L}$ when a droplet reaches its maximum spreading factor, (b) circularity of the triple contact line $\delta$ when a droplet reaches its maximum spreading factor.}
  \hfill
\end{figure}

Figure~\ref{pD_CLLenngth} shows the dimensionless contact line length $\tilde{L}$ when droplets reach their maximum spreading factors. The dimensionless contact line length is defined by $\tilde{L}=L_c/D_0$ with $L_c$ and $D_0$ being the length of triple contact line and droplet diameter, respectively. 
At each fixed $R_q$, Figure~\ref{pD_CLLenngth} evidences that increasing \textit{We} leads to a larger normalized contact line length $\tilde{L}$. Evidently, $\tilde{L}$ increases from 5.22 (\textit{We}=5.7) to 5.97 (\textit{We}=12.9) when $R_q$ = 20 $\mu m$. 
For each given \textit{We}, $\tilde{L}$ gradually increases as the surface roughness $R_q$ increases. For instance, $\tilde{L}$ rises from around 5.1 ($R_q=0$) to approximately 5.39 ($R_q=50$) at \textit{We}=5.7. Note that $\tilde{L}$ shows slightly decrease as $R_q$ increases for a given \textit{We}, most noticeable at low $R_q$ region ($R_q \leq 10 \ \mu m$). As indicated in Figure~\ref{maxSpreadF}, the maximum spreading factor $\beta_m$ shows a minor reduction as $R_q$ increases from 0 to 10 $\mu m$. The triple contact line on a smooth surface appears as a perfect two-dimensional circle when viewed from above. The rough surfaces are approximately smooth when their $R_q \leq 10 \ \mu m$, and thus triple contact lines on them are nearly two-dimensional curves. The variation of $\tilde{L}$ in the direction parallel to the surface normal is negligible. Therefore, its contribution to the total contact line length can be neglected. In contrast to surfaces with smaller roughness values ($R_q \leq 10 \ \mu m$), triple contact lines on rough surfaces with $R_q > 10 \ \mu m$ form three-dimensional curves, and thus a dramatic variation of $\tilde{L}$ along the roughness height direction cannot be neglected. Increased surface roughness enhances the contribution of $\tilde{L}$ from the roughness height direction, thereby leading to larger $\tilde{L}$ values on rough surfaces. 

In order to quantify the evolution of the three-dimensional triple contact line on rough surfaces, we define a dimensionless circularity given by $\delta={4 \pi A_t}/{L_c^2}$ with $A_t$ being the total contact area between the droplet and the surface. Figure~\ref{pD_circularity} shows $\delta$ at different \textit{We} and $R_q$ when droplets reach their maximum spreading factors. Clearly, $\delta$ is approximately 1 ($> 0.999$) on smooth surfaces ($R_q$=0) across all five Weber numbers, indicating almost perfect circular contact lines. For each given \textit{We}, a pronounced decrease in contact line circularity is observed with increasing surface roughness. For instance, $\delta$ drops from 0.999 ($R_q$ = 0) to 0.935 ($R_q$ = 20) and further to 0.888 ($R_q$ = 50) at \textit{We} = 5.7. $\delta$ varies between 0.822 and 0.888 on the roughest surfaces ($R_q$ = 50), suggesting a highly irregular contact line shape, as evidenced by its strong deviation from the circularity of a prefect circular contact line. Note that the circularity of the triple contact line on smooth surfaces is independent of Weber number when $5.7 \leq$ \textit{We} $\leq 12.9$; however, $\delta$ depends on both surface roughness and Weber number on rough surfaces, especially when $R_q \geq 20$. Furthermore, a comparative analysis across different Weber numbers evidences that, for a given $R_q$, an increase in \textit{We} correlates with a slight further reduction in circularity. However, the magnitude of circularity degradation caused by increasing roughness far exceeds that associated with increasing Weber number. This indicates that surface morphology plays a more dominant role than Weber number in determining the resultant circularity.

\subsubsection{Wetting dynamics}
\begin{figure}[htp]
  \begin{center}
    \includegraphics[width=0.5\textwidth]{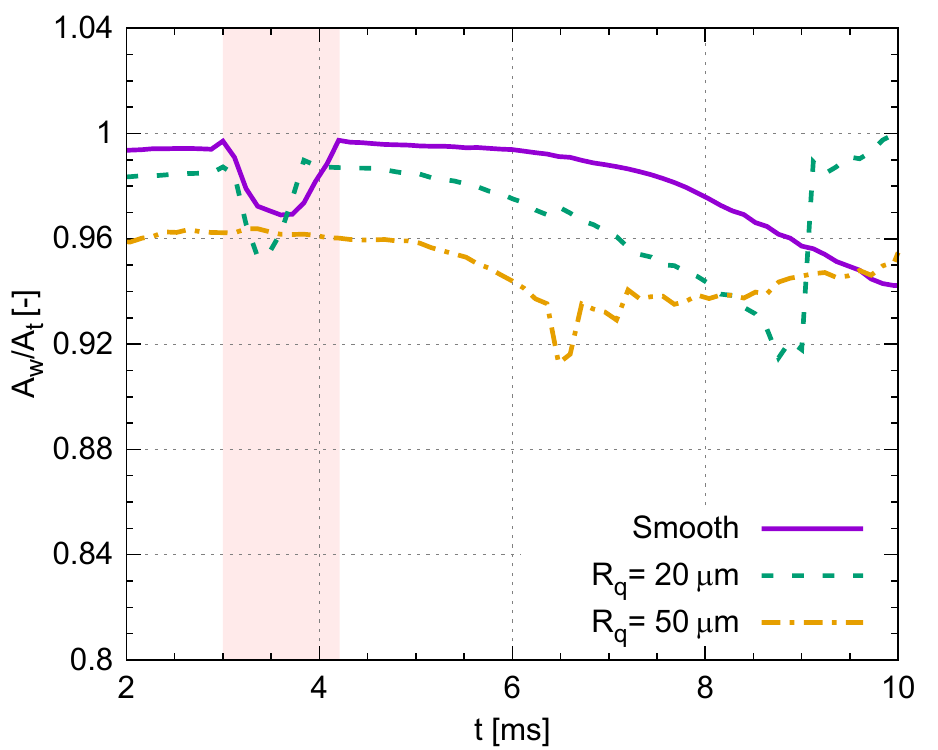}
  \end{center}
  \caption{Temporal evolution of $A_w/A_t$ at \textit{We}=5.7. The pink shaded band indicates the region where droplets reach their maximum spreading factors.}
  \label{wettingAreaRatio}
\end{figure}

Some air bubbles are entrapped within droplets during droplet impact, and we denote the area of the surface covered by bubbles as $A_b$. Thus, the total droplet contact area on surfaces $A_t$ is calculated by $A_t=A_w+A_b$ with $A_w$ being the droplet wetting area. Figure~\ref{wettingAreaRatio} shows the temporal evolution of $A_w/A_t$ when \textit{We}=5.7.
A value of $A_w/A_t = 1$ corresponds to complete wetting, whereas $A_w/A_t < 1$ indicates partial wetting with part of the surface is occupied by bubbles. As surface roughness $R_q$ increases from 0 to 20 and then to 50, the ratio $A_w/A_t$ progressively decreases from approximately 0.993 to 0.982 and finally to 0.961 at $t$ = 2 \si{ms}. This demonstrates that surface morphology promotes air entrapment, which subsequently shifts the system from the Wenzel state immediately after contact. A sharp reduction in $A_w/A_t$ is observed during the spreading phase, corresponding to a concurrent, gradual increase in $A_b$. Similarly, droplet impact on a surface with $R_q$ = 20 also exhibits a reduction. In contrast to the above two cases, a decrease in $A_w/A_t$ occurs during the recoiling phase when $R_q$ = 50. Overall, surface roughness enhances air entrapment during both spreading and receding phases, thereby critically influencing the wetting dynamics.

\subsubsection{Internal flow dynamics}
\begin{figure}[htp]
  \begin{center}
    \includegraphics[width=0.95\textwidth]{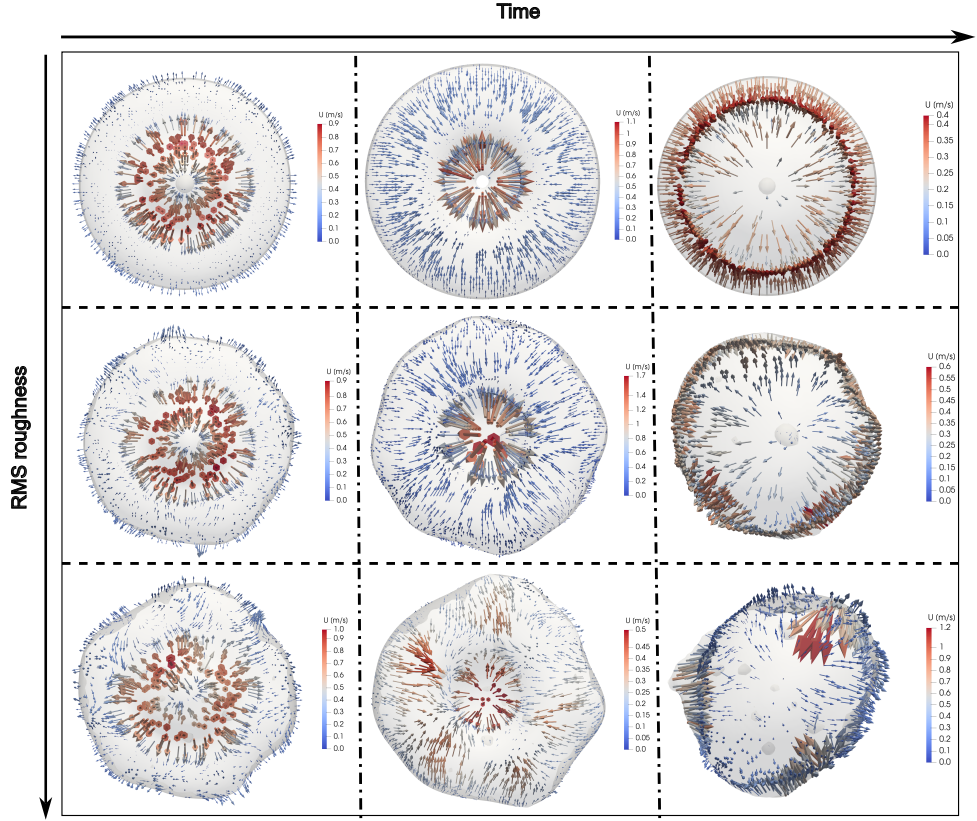}
  \end{center}
  \caption{Internal flow dynamics and velocity vectors inside impacting droplet at \textit{We}=5.7 (top view). Top row: smooth surface, middle row: $R_q$=20 $\mu m$, bottom row: $R_q$=50 $\mu m$; left column: during spreading phase, middle column: at the maximum spreading factor, right column: during the receding phase.}
  \label{velocityVector}
\end{figure}

Figure~\ref{velocityVector} shows internal velocity vectors and droplet deformation during both spreading and receding phases at \textit{We}=5.7 when viewed above. 
The top, middle, and bottom rows demonstrate droplet deformation and velocity vectors for impact on surfaces with $R_q$ = 0, 20, and 50 $\mu m$, respectively. Each column corresponds to a distinct phase (left to right): velocity vectors and droplet morphology during the spreading phase, at the maximum spreading factor, and during the receding phase. In addition to the circularity information presented in Figure~\ref{pD_circularity}, we can clearly observe the effect of surface roughness on droplet deformation from Figure~\ref{velocityVector}. As surface roughness $R_q$ increases from 0 to 20 and further to 50 $\mu m$, droplets at three different stages become increasingly irregular, accompanied by a gradual loss of geometric symmetry. Furthermore, velocity vectors inside droplets also demonstrate roughness-dependent features. On smooth surfaces, both internal flow and velocity vectors around the triple contact line exhibit strong symmetry and thus droplets can maintain its circularity during impact. As indicated by the sub-figure on the top right, the triple contact line retracts inward at a constant velocity, and meanwhile, the droplet center remains stationary. In contrast to smooth surfaces, velocity fields adjacent to triple contact lines exhibit disordered asymmetry in both magnitude and direction when impact on rough surfaces, and this asymmetry is enhanced by increased surface roughness. As discussed before, surface morphology affects the shape of the triple contact line. Additionally, increased surface roughness significantly contributes to the variation in the roughness height direction. In addition to its effect on flow dynamics around the triple contact line, Figure~\ref{velocityVector} further evidences that random surface morphology also significantly alters internal flow. To be specific, the middle column demonstrates that increased surface roughness enhances the internal flow as $R_q$ rises from 20 to 50 $\mu m$. 

\section{Conclusions}
\label{conclusion}
In this work, volume of fluid simulations have been employed to understand how surface roughness affects droplet impact dynamics on hydrophobic surfaces with an equilibrium contact angle of $100^{\circ}$. By varying the root-mean-square roughness $R_q$ and Weber number \textit{We}, we examined 35 numerical cases. The impact outcomes of the 35 cases have been classified into three categories: no bouncing, complete bouncing, and bouncing with breakup. The main conclusions and contributions are summarized as follows:

(1) Droplets spread more slowly on rough surfaces than on smooth ones, as evidenced by the temporal evolution of spreading factors. Re-spreading appears during the receding phase for droplet impact on rough surfaces with $R_q \geq 30 \ \mu m$ at each given Weber number ranging from 5.7 to 12.9. The maximum spreading factors $\beta_m$ exhibit approximately linear reduction as $R_q$ increases from 0 to 50 $\mu m$. $\beta_m=1.14(1.38+\textit{We})^{0.46}$ and $\beta_m=1.11(1.55+\textit{We})^{0.40}$ have been proposed for slightly rough surfaces ($R_q \leq 20 \ \mu m$) and rough surfaces ($R_q \geq 30 \ \mu m$), respectively. 

(2) Surprisingly, droplet contact time $\tau_c$ on hydrophobic surfaces has been proved to be independent of both Weber number and RMS roughness, and $\tau_c=3.9\sqrt{\rho R^3/\sigma}$ is herein proposed in this work. 

(3) The wetting-bouncing transition is triggered by increasing \textit{We} and decreasing $R_q$. Droplet impact on rough surfaces ($R_q > 10 \ \mu m$) results in larger triple contact line length as compared to impact on smooth surfaces. On smooth surfaces, droplets manage to maintain their spherical triple contact line, whereas on rough surfaces, the triple contact line becomes more irregular, as evidenced by decreased circularities. 

(4) Additionally, as $R_q$ increases, more air bubbles are entrapped inside droplets and thus significantly affect droplet wetting dynamics. Visualizations of internal flow and velocity vectors during droplet impact uncover that the velocity field around triple contact line shows disordered asymmetry when impact on rough surfaces. Random surface morphology significantly alters internal flow dynamics and induces instabilities, especially when impact on surfaces with larger $R_q$. 

The wetting-bouncing transition identifies surface morphology as another dominant factor governing droplet impact outcomes, providing a simple yet efficient way to manipulate droplet impact modes by tailoring surface morphologies. The linear scaling of maximum spreading factor with RMS roughness serves as a simple predictive tool for droplet spreading dynamics on both ideally smooth surfaces and real-world rough substrates. The invariance of contact time with impact conditions offers a robust design principle for controlling droplet-surface interactions, particularly for developing water-repellent and anti-icing surfaces, optimizing spray coating, and regulating agricultural spraying. Moreover, by elucidating the mechanistic link between surface morphology, internal flow dynamics, and macroscopic impact behavior, this work advances a foundational framework for the optimal design of surfaces with tailored wetting and bouncing characteristics, relevant to a wide range of colloidal and interfacial applications.

In future work, droplet impact on superhydrophobic surfaces will be of particular interest. Special attention will be given to droplet dynamics on moving surfaces, whether in translational or rotational motion.

\section *{Acknowledgements}
This work was supported by the Major Program of National Natural Science Foundation of China (Grant No. 22293024) and the Strategic Priority Research Program of the Chinese Academy of Sciences (Grant No. XDA0390501). This work was carried out using the facilities at Huairou Interdisciplinary Research Center for Mesoscience, CAS-IPE. We thank Chunxiao Zhu (Sichuan University) and Mian Zeng (University of Chinese Academy of Sciences) for their kind help in data post-processing using ParaView.

\section *{Appendix A:~Numerical setup}
\label{modelSetup}
The numerical setup is shown in Figure~\ref{modelSetup}, and a droplet with a diameter $D_0 = 1.6 \ \si{mm}$ is initialized above the rough surface. 
\begin{figure}[htp]
  \begin{center}
    \includegraphics[width=0.5\textwidth]{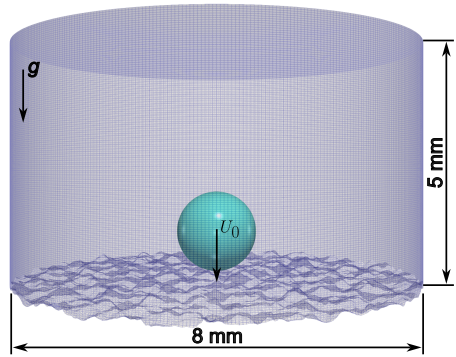}
  \end{center}
  \caption{Schematic diagram of the numerical setup.}
  \label{modelSetup}
\end{figure}
A cylindrical computational domain with a rough surface on the bottom is used in our numerical simulations, and the diameter and height of the computational domain are $8 \ \si{mm}$ and $5 \ \si{mm}$, respectively. Note that when simulating droplet impact on smooth surfaces, the roughness of the computational domain is zero. The initial separation distance between the droplet’s lowest point and the base plane of the rough substrate is $0.1 \ \si{mm}$. The droplet moves towards the bottom rough/smooth wall with an initial velocity $U_0$. $\textbf{\textit{g}}$=9.81 [\si{m/s^2}] is the gravitational acceleration in the vertical direction. The grid size $\Delta x$ is equal to $D_0/50$, and the validity of such a mesh configuration has been demonstrated in the literature~\cite{amani2024direct}. Essential parameters of both phases are listed in Table~\ref{KeyParameters}.
\begin{table}[htp]
\caption{Parameters used in droplet impact simulations.}
\label{KeyParameters}
\centering
\begin{tabular}{cccc}
\hline
Phase & Density $[\si{kg/m^3}]$ & Dynamic viscosity $[\si{Pa \cdot s}]$ & Surface tension $[\si{N/m}]$ \\ \hline
Liquid & 1000    & $1\times 10^{-3}$               & 0.07    \\
Gas   & 1       & $1.48\times 10^{-5}$            & -  \\ \hline
\end{tabular}
\end{table} 
The contact angle boundary condition with an equilibrium contact angle of $100^{\circ}$ is applied on the bottom rough/smooth surface. The pressure outlet boundary conditions are applied on the other boundaries, except for the bottom wall, where a no-slip boundary condition is prescribed. The total simulation time is $12 \ \si{ms}$, and it guarantees impacting droplets experience spreading, recoiling, and rebound phases. 

\section *{Appendix B:~Model validation}
\label{modelValidation}
In this section, we compare our numerical results against experimental and numerical data adopted from the literature~\cite{rioboo2002time,amani2024direct}.  A droplet with a diameter of $2.75 \ \si{mm}$ moves towards a smooth surface with an initial velocity of $1.18 \ \si{m/s}$. Similar to the numerical setup shown in Figure~\ref{modelSetup}, a cylindrical computational domain with a radius of $5.5 \ \si{mm}$ and a height of $11 \ \si{mm}$ is used for numerical validations in this section. The initial distance between the droplet center and the bottom surface is $1.375 \ \si{mm}$. An equilibrium contact angle $\theta_e=99.76^{\circ}$ is prescribed on the bottom surface. Additionally, we prescribe the no-slip boundary condition on the bottom surface and the pressure outlet boundary condition on the other surfaces. 
\begin{figure}[htp]
  \begin{center}
    \includegraphics[width=0.5\textwidth]{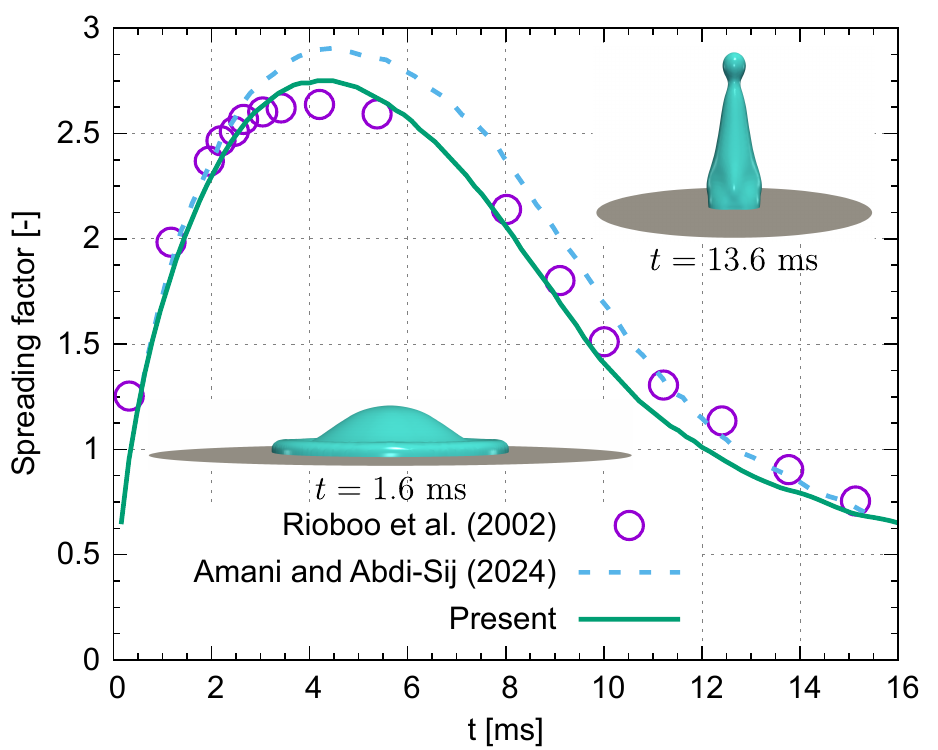}
  \end{center}
  \caption{Evolution of the spreading factor versus time. Two insets show droplet deformation during the spreading (at $t=1.6$ \si{ms}) and recoiling (at $t=13.6$ \si{ms}) phases, respectively.}
  \label{SeeberghModel}
\end{figure}
Here, we define the dimensionless spreading factor as the ratio of the wetting diameter to the initial droplet diameter before impact. As shown in Figure~\ref{SeeberghModel}, purple circles represent experimental results from~\cite{rioboo2002time}, and the dashed line shows numerical results presented in the literature~\cite{amani2024direct}. The solid line shows our numerical results for predicting the spreading factor during droplet impact. The droplet starts to spread once it contacts with the bottom surface, reaching its maximum spreading factor at $t\approx4.2 \ \si{ms}$, followed by the recoiling phase. Two insets of Figure~\ref{SeeberghModel} present snapshots of the spreading and receding phases, respectively. Figure~\ref{SeeberghModel} demonstrates good agreement between our numerical results and the experimental data. Note that our numerical model shows better performance in predicting the spreading factor than the model presented in the literature~\cite{amani2024direct}, which overestimates the maximum spreading factor. 

To validate the performance of the W-M function for generating random rough surfaces, three random surfaces with $R_q$= 50 $\mu m$ were generated. Droplet impacting on the above three surfaces with \textit{We}=5.7 was conducted.
\begin{figure}[htp]
  \begin{center}
    \includegraphics[width=0.5\textwidth]{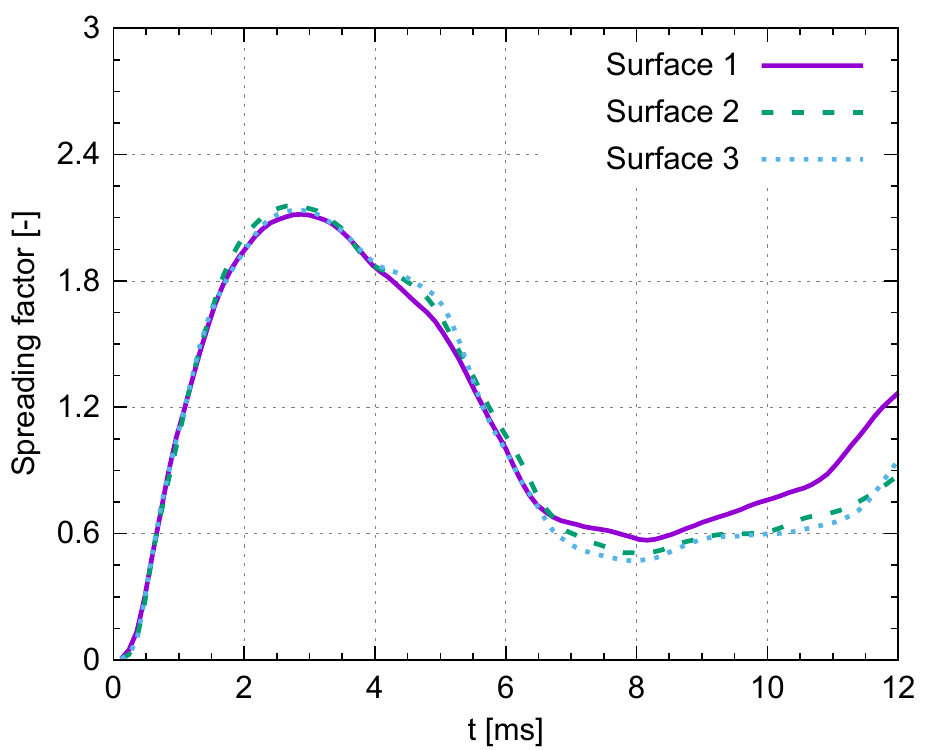}
  \end{center}
  \caption{Temporal evolution of the spreading factor for three random surfaces.}
  \label{Rq50Comparison}
\end{figure}
Figure~\ref{Rq50Comparison} shows the temporal evolution of the spreading factor, and three surfaces demonstrate negligible discrepancy during the spreading phase and the early stage of the receding phase. Minor discrepancies are observed during the late retraction and re-spreading phases. During the spreading and early retraction stages, the liquid film is much thicker than the characteristic roughness height, and thus the flow is dominated by inertia and capillary forces rather than surface morphology. In contrast, during the late retraction and re-spreading phases, the film becomes significantly thinner, even comparable to or smaller than the roughness height. Consequently, the droplet dynamics become particularly sensitive to local contact line dynamics governed by the random fractal surface morphology.

\section *{Appendix C:~Maximum spreading factor $\beta_m$ as a function of \textit{We/Oh}}
\label{appendixC}
Similar to the scaling law shown in Figure~\ref{scalingLaw_We}, we also identify two regimes governed by surface roughness values for the correlation between $\beta_m$ and \textit{We/Oh}.
\begin{figure}[htp]
  \begin{center}
    \includegraphics[width=0.5\textwidth]{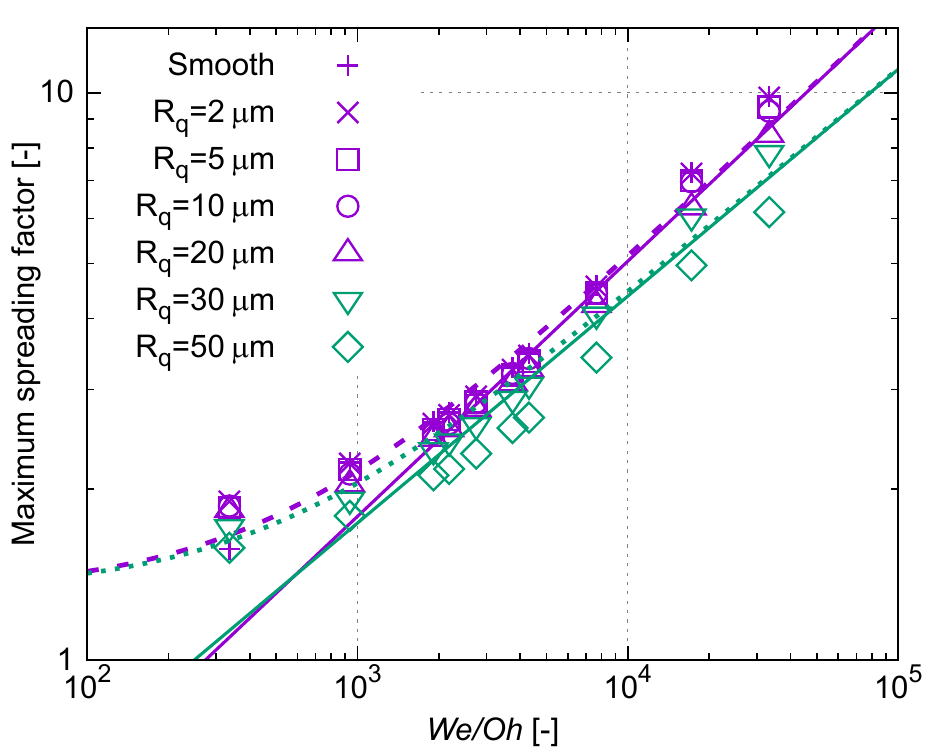}
  \end{center}
  \caption{Maximum spreading factor $\beta_m$ as a function of \textit{We/Oh}. Solid purple and green lines represent $\beta_m=0.08(\textit{We/Oh})^{0.45}$ and $\beta_m=0.11(\textit{We/Oh})^{0.40}$, respectively. Dashed purple and green lines represent $\beta_m=0.08(509.4+\textit{We/Oh})^{0.45}$ and $\beta_m=0.11(500.6+\textit{We/Oh})^{0.40}$, respectively.}
  \label{scalingLaw_WeOh}
\end{figure}
For surfaces with lower roughness ($R_q \leq 20 \ \mu m$), $\beta_m$ follows $\beta_m=0.08(509.4+\textit{We/Oh})^{0.45}$. $\beta_m=0.11(500.6+\textit{We/Oh})^{0.40}$ is found for surfaces with higher roughness ($R_q \geq 30 \ \mu m$). As indicated in Figure~\ref{scalingLaw_WeOh}, the semi-empirical scaling model $\beta_m \sim (509.4 + \textit{We/Oh})^{0.45}$ shows better performance in predicting $\beta_m$ than $\beta_m \sim (\textit{We/Oh})^{0.45}$.
\end{spacing}

\bibliographystyle{elsarticle-num}
\bibliography{Refs}
\end{document}